# Artificial intelligence and cybersecurity in banking sector: opportunities and risks

Ana Kovačević [1]

Sonja D. Radenković [2]

Dragana Nikolić [3]

*Abstract*: The rapid advancements in artificial intelligence (AI) have presented new opportunities for enhancing efficiency and economic competitiveness across various industries, espcially in banking. Machine learning (ML), as a subset of artificial intelligence, enables systems to adapt and learn from vast datasets, revolutionizing decision-making processes, fraud detection, and customer service automation. However, these innovations also introduce new challenges, particularly in the realm of cybersecurity. Adversarial attacks, such as data poisoning and evasion attacks, represent critical threats to machine learning models, exploiting vulnerabilities to manipulate outcomes or compromise sensitive information. Furthermore, this study highlights the dual-use nature of AI tools, which can be used by malicious user. To address these challenges, the paper emphasizes the importance of developing machine learning models with key characteristics such as security, trust, resilience and robustness. These features are essential to mitigating risks and ensuring secure deployment of AI technologies in banking sectors, where the protection of financial data is paramount. The findings underscore the urgent need for enhanced cybersecurity frameworks and continuous improvements in defensive mechanisms. By exploring both opportunities and risks, this paper aims to guide the responsible integration of AI in the banking sector, paving the way for innovation while safeguarding against emerging threats.

*Keywords*: Artificial Intelligence, Machine Learning, Cyber Security, Adversarial Attacks, Banking

## 1. INTRODUCTION

Artificial intelligence is reshaping the banking industry, offering unparalleled opportunities for innovation and efficiency. According to Gartner (2024b), 50% of respondents in the banking sector reported having already implemented AI, while the adoption of generative AI, although growing, remains less common (40%). Furthermore, the prediction is that by 2027, over 50% of enterprises will employ industry-specific generative AI models, a significant increase from the 1% that utilised such models in 2023 (Sau et al., 2024).

[1] Faculty of Security Studies, University of Belgrade, kana@fb.bg.ac.rs, https://orcid.org/0000-0003-4928-9848
[2] Belgrade Banking Academy – Faculy of Banking, Insurance and Finance, Union University Belgrade, sonja.radenkovic@bba.edu.rs, https://orcid.org/0000-0001-6830-7533

[3] Institute of Nuclear Sciences Vinca, University of Belgrade, anikol@vinca.rs, https://orcid.org/0000-0003-2568-7729

Generative Artificial Intelligence (Generative AI) refers to large-scale models trained on billions of parameters to generate diverse media types. However, this transformative technology also introduces new security challenges, making cybersecurity a very important priority. Organisations plan to significantly increase their investments in 2025 compared to 2024, prioritising cybersecurity (89%), generative AI (90%), and broader AI applications (85%) (Gartner, 2024b). Overall, the report emphasises a clear trend: organisations in the banking industry recognise the importance of adopting advanced technologies on the need to innovate services, and stay competitive while also bolstering security measures to address an increasingly complex threat landscape. Cybersecurity remains one of the top priorities as the sector seeks to protect sensitive financial data, prevent cyber attacks, and ensure compliance with stringent regulatory requirements. This underscores the urgent need for robust security measures to address an increasingly complex threat landscape.

One of the biggest reasons the banking sector is usually targeted by most cyber-attacks is the monetary gain (Chin, 2024). Given that financial data is the core of the banking sector, any attack on the system might incapacitate a company and decrease customer trust in the company. In 2024, Kaspersky (2024) data revealed a significant global rise in mobile financial threats, with the number of affected users doubling compared to 2023. Further, more upward trend will persist into 2025.

## 2. APPLICATION OF AI IN THE BANKING SECTOR

The competitive environment in the banking sector is progressively influenced by digital-first entities, including fintech companies and digital banks, rather than conventional banks with significant branch networks. This transition underscores the necessity to examine the impact of artificial intelligence (AI) on the banking industry, especially in facilitating digital transformation (Aithal, 2023). AI has become a crucial technology transforming banking operations via new solutions that improve scalability, efficiency, and customer experience.

Integrating AI in banking centres on four key digital disruptors: advanced data analytics, robots, embedded banking, and intelligent infrastructure. These improvements facilitate several uses, including as AI-driven chatbots for customer service, robo-advisors for investment advice, predictive analytics, improved cybersecurity, automated credit assessment, and direct financing.

Implementing AI in digital banking can be classified into three sequential tiers (Radenković et al., 2023). The initial stage entails utilizing machine learning to discern patterns and enhance digital banking operations. The second level presents "general intelligence" that can replicate human interactions so successfully that customers and workers may be unaware they are engaging with a machine. The third and most sophisticated level includes AI systems that exceed human banking personnel in intellect and decision-making capabilities.

The integration of AI has significant advantages, although it is a complicated, protracted process necessitating the satisfaction of multiple operational and regulatory requirements. The emergence of AI in banking also induces considerable labour transformations. Conventional roles are anticipated to diminish while emerging positions—such as data scientists, behavioural psychologists, and experience designers—are increasingly crucial for enabling the shift to AI-driven digital banking. This trend highlights the dual impact of AI: it undermines the existing paradigm while fostering chances for innovation and growth in the financial sector.

## 3. CYBER SECURITY

Cybersecurity is a complex, computing-based discipline integrating technology, people, information, and processes to protect systems from unauthorised access or attacks (Joint Task Force on Cybersecurity, 2018). However, growing concerns exist about the potential risks of deploying AI in security-critical areas. Security remains a top priority for corporate boards when considering AI adoption (Gartner, 2024a). The risks associated with AI failures or misuses are real and could lead to severe consequences, particularly in sensitive sectors such as banking. These concerns are particularly pressing given the current environment, where cyber attacks are growing in both frequency and sophistication, leading to significant costs for organisations in defence measures and mitigation of potential harm. Preventing such attacks is especially important, as their disruption can profoundly affect security and functionality of banking sectors.

With the rapid advancements in artificial intelligence (AI), a pressing question emerges: will AI serve as a tool to address cybersecurity challenges, or will it make (generate) new, sophisticated attack vectors? The rise of generative AI has already showcased its dual-use potential, offering both defensive capabilities and opportunities for malicious exploitation. Hazell (2023) conducted a hypothetical experiment to demonstrate how generative AI like GPT-3 and GPT-4 could automate the creation of spear-phishing campaigns. The process involved gathering personalised biographical data from online sources, crafting customised email messages based on this data, and embedding malware within those emails. This highlights a critical concern: how will malware and cyber attacks evolve with AI support? Will these tools enable attackers to launch more dangerous and sophisticated campaigns? WormGPT exemplifies this threat. Built on the open-source GPT-J framework, WormGPT has been trained specifically on data related to malware and phishing emails. Unlike ethical models like ChatGPT, WormGPT operates without safeguards, enabling the efficient generation of malicious content. Beyond WormGPT, other tools such as AutoGPT, ChatGPT with DAN prompts, FreedomGPT, FraudGPT have emerged, each offering various capabilities to generate harmful content . These developments underscore the inherent dual-use nature of AI technologies, highlighting the urgent need for comprehensive regulatory frameworks and ethical guidelines to mitigate their potential misuse.

Moreover, research conducted by Oracle &KMPG (2019) highlights a significant shortage of cybersecurity experts. Given the rise in cyberattacks and the shortage of skilled personnel, there is a growing need to automate threat detection and response processes where AI can offer significant support (Kocher and Kumar 2021; Kovačević 2023).

## 4. ARTIFICIAL INTELIGENCE

With advancements in hardware and software and the generation of large amounts of data, artificial intelligence has been experiencing rapid expansion. Machine learning is a subset of artificial intelligence. It refers to a software system's ability to generalise based on prior experience, where experience is defined as a dataset concerning the phenomena/entities that are the subject of learning.

 Based on the type of decision-making process, machine learning can be categorised as supervised learning, unsupervised and reinforcement learning (Kovačević 2023). In cases where humans have achieved better results, machine learning has been observed to improve accuracy with increased training, which is not the case for humans. It has proven highly effective in analysing large volumes of data and detecting previously unknown patterns, as well as making it helpful in detecting and blocking cyber attacks (Kovačević 2023), such as identifying and preventing system intrusions or detecting malicious code. Additionally, it can investigate attacks on targeted systems, analyse entry possibilities, and uncover vulnerabilities. Machine learning can also detect unusual activities that indicate cyber attacks or system misuse.

## 4.1. Machine learning in analyzing cyber security attacks

Detecting cyber attacks remains a substantial challenge, often requiring months or even years to identify an intrusion (Sobers, 2022). Given the vast and growing data volume in modern computing environments, machine learning offers a powerful approach to improve cyber attack detection. Zero-day vulnerabilities, in particular, are increasingly central to advanced cyber operations; leveraging machine learning to detect these new vulnerabilities could significantly enhance cybersecurity defences.

Machine learning systems are particularly valuable for identifying software vulnerabilities within complex codebases, which can contain millions of lines of code. This complexity makes manual analysis time-consuming and prone to human error. By training machine learning models to recognise these vulnerabilities, the process can be effectively automated, enabling faster and more accurate detection. As machine learning systems continuously improve through experience and training data, the potential for precise increasing vulnerability detection grows (Saavedra et al., 2019).

The application of machine learning is equally promising in developing advanced intrusion detection systems (IDSs). Different machine learning methodologies - supervised, unsupervised, and reinforcement learning - bring distinct advantages to IDSs. Supervised learning effectively identifies well-documented attacks with low false favourable positive rates, though it struggles with unknown threats (Shaukat et al., 2020; Pinto et al., 2023). Unsupervised learning holds the potential for detecting zero-day attacks; however, it often results in a higher rate of false positives (Almalawi et al., 2020). Reinforcement learning, if given sufficient training time, has the potential to adapt dynamically to evolving cyber threats, providing a robust defence (Nguyen & Reddi, 2023). The integration of these advanced techniques has led to notable improvements in the accuracy and responsiveness of IDSs, further advancing the field of cybersecurity.

## 4.2. Attacks on machine learning models

The development of machine learning models requires the collection of large amounts of data. It can and can be very costly, mainly due to the time needed for the development, tuning, and validation of the models. Attackers may exploit vulnerabilities in these models to gain unauthorised access to sensitive information. Attacks on machine learning can also significantly reduce the effectiveness of these technologies, leading to devaluation of the investments made in them and delayed implementation (Koball et al., 2024).

By analysing threat vectors in the context of attacks on machine learning, the following types of attacks are identified (Koball et al., 2024):

- Data Extraction: The attacker attempts to uncover the data on which the model was trained, which is particularly dangerous if the model uses sensitive data.
- Data Poisoning: The attacker manipulates the training data, introducing false examples to make the model produce inaccurate predictions. Due to the inherent nature of the model, minimal changes in input data, imperceptible to humans, can lead to misclassification.

- Model Extraction: The attacker obtains information about the internal structure of the model.
- Evasion attack: The attacker forces the model to make incorrect predictions and avoids detection by generating adversarial examples.
- Data poisoning and evasion attacks influence machine learning output, while data extraction and model extraction are passive attacks. An example of a data poisoning attack occurred in a financial institution where attackers manipulated training data for fraud detection, resulting in the system approving fraudulent transactions as legitimate.

We can consider a hypothetical scenario in which a machine learning model detecting security incidents, such as unauthorised access or breaches. In this scenario, attacker could insert manipulated data into the training set. This could alter the model's ability to detect security incidents accurately, potentially allowing unauthorised individuals or malicious activities to go unnoticed. Furthermore, an example of a data poisoning attack occurred in a financial institution where attackers manipulated training data for fraud detection, resulting in the system approving fraudulent transactions as legitimate.

The widespread use of machine learning algorithms has led to malicious manipulations known as adversarial attacks. These attacks impact the decision-making process in machine learning, resulting in outcomes that benefit the attacker (Goodfellow et al., 2014a; Goodfellow *et al.,* 2014b). The way neural networks perform data classification differs from human perception, so even a small change in just a few pixels can significantly alter the classification result, even if a human cannot perceive the difference (Goodfellow et al., 2014b). Additionally, a change in the position of an object can lead to misclassification (Alcorn et al., 2019), while Li et al., (2019) demonstrated how a facial recognition system can produce incorrect results even with minimal object shifts. Based on the level of access to the machine learning model, attackers can target systems in the following categories (Kobal et al., 2024):

- *Black Box*: the attacker does not have insight into the model's details but uses input and output data to analyse vulnerabilities.
- *White Box:* the attacker has complete knowledge of the model's architecture, enabling them to create precise perturbations to exploit its weaknesses.
- *Gray Box*: these attacks grant the attacker partial access to information, somewhere between full access in white-box attacks and the limited data available in black-box attacks.

On the other hand, if a machine learning model is trained on confidential data, model inversion can allow an attacker to reveal key characteristics of the underlying data used for training the system (Buchanan, 2021).

A machine learning model must possess several key characteristics, especially in areas of great importance to society: security, trust in the model, robustness, and resilience (Khadka et al., 2024). Robustness refers to the model's ability to function even during crises, while resilience is the system's ability to return to normal functioning within a reasonable time after a disruption. These characteristics are crucial in demanding domains, as their absence can have serious consequences, particularly in critical infrastructures. Achieving resilience and robustness can significantly improve trust and security within the system.

To achieve security, trust in the model, resilience, and robustness, various strategies can be applied, as suggested by Khadka et al., (Khadka et al., 2024), such as: understanding and simulating different types of attacks, detecting adversarial attacks, training robust models and/or understanding the weaknesses and vulnerabilities of the system.

Some defensive strategies are implemented during the training phase, while others are used during the testing phase. Adding adversarial data to the training set can help the model become more resilient to adversarial perturbations during training. Based on the available literature, Khadka et al. (2024) mention some of the most commonly used strategies for adversarial training, such as brute-force adversarial training, data randomisation, gradient masking, and more.

Machine learning security is crucial as machine learning-assisted methods become more widespread. Research indicates a significant gap in understanding the full extent of attack success rates against these systems. Addressing this gap is essential to developing robust defences and improving the resilience of machine learning models in practical applications.

## 5. CONCLUSION

As a core element of modern artificial intelligence systems, machine learning finds broad application in banking, delivering numerous benefits, such as enhanced efficiency, improved decision-making, trend prediction, anomaly detection, and security monitoring. However, one evident trend is that machine learning's integration will add complexity to traditional attack vectors, increasing the risk of cyber operations and reshaping the nature of cyber threats. The expansion of machine learning introduces specific security challenges, such as adversarial attacks or data/system extraction. These threats, posed through white, grey, or black box methods, seriously endanger system privacy and data integrity. As a result, it is imperative to assess machine learning technologies thoroughly within sensitive systems, recognising that they come with distinctive risks and advantages. Ongoing research into potential attack methods and their effects on machine learning models is critical.

This research must include developing defence strategies encompassing vulnerability assessments, attack simulations, and training of robust models. The resilience and robustness of machine learning models, key traits enabling them to function effectively under threat, are essential to building trust in these technologies. Continuous improvement of defensive mechanisms through research and testing is needed to ensure the secure use of machine learning in the banking sector. It is essential to carefully evaluate the deployment of machine learning in sensitive environments like banking, where unique risks tied to this technology may also emerge. While machine learning is not inherently more dangerous than human action, it can operate at computational speeds and scales beyond human capacity, amplifying its benefits and potential risks. Banking institutions should prioritize investment in adversarial training and robust model validation to mitigate risks associated with AI vulnerabilities.

In particular, ensuring generative AI systems' security and robustness is vital to responsibly leveraging their capabilities the security and robustness of generative AI systems is vital to leveraging their capabilities responsibly, especially in critical areas with significant societal impact. Machine learning holds considerable promise in vulnerability detection and cyber attack prevention. However, crucial questions remain: will machine learning accelerate or resolve cyber attacks? Will attackers benefit more from its application, or will it fortify defences? Achieving a comprehensive understanding of machine learning's capabilities and limitations is crucial to mitigating misuse. Future research should focus on the regulatory framework for AI use in critical sectors, ensuring a balance between innovation and security.

**ACKNOWLEDGEMENT**

***This work was supported by the Science Fund of the Republic of Serbia under Grant 7749151 within the Framework of the IDEAS Program—Management of New Security Risks Research and Simulation Development, NEWSiMR&D.***